\documentclass[a4paper]{jpconf}
\usepackage{graphicx}
\usepackage{amsmath}
\usepackage{subcaption}
\usepackage{siunitx}					% For the unit command \SI{3.25(2)}{\kilo\gram} 
\usepackage[colorlinks=true, a4paper=true, pdfstartview=FitV,
linkcolor=black, citecolor=black, urlcolor=black]{hyperref}
\usepackage{url}

\usepackage{graphicx}% Include figure files
\usepackage{dcolumn}% Align table columns on decimal point
\usepackage{bm}% bold math
\usepackage[USenglish]{babel}
\usepackage[separate-uncertainty]{siunitx}% Nice micron signs and units
\usepackage{multirow}
\usepackage{textcomp}
\usepackage{hhline}
\usepackage{tabularx}
\usepackage[flushleft]{threeparttable}
\usepackage[table]{xcolor}
\usepackage{comment}
\usepackage{cite}

\usepackage{scalerel}

\DeclareFontFamily{U}{BOONDOX-calo}{\skewchar\font=45 }
\DeclareFontShape{U}{BOONDOX-calo}{m}{n}{
  <-> s*[1.05] BOONDOX-r-calo}{}
\DeclareFontShape{U}{BOONDOX-calo}{b}{n}{
  <-> s*[1.05] BOONDOX-b-calo}{}
\DeclareMathAlphabet{\mathcalboondox}{U}{BOONDOX-calo}{m}{n}
\SetMathAlphabet{\mathcalboondox}{bold}{U}{BOONDOX-calo}{b}{n}
\DeclareMathAlphabet{\mathbcalboondox}{U}{BOONDOX-calo}{b}{n}

\begin{document}

%\title{Status of, and upgrade concepts for, the hybrid, asymmetric, linear Higgs factory (HALHF)}
\title{Status of and upgrade concepts for HALHF: \\the hybrid, asymmetric, linear Higgs factory}

\thispagestyle{plain}
\pagestyle{plain}

\author{C A Lindstrøm\textsuperscript{1}, R D'Arcy\textsuperscript{2,3} and B Foster\textsuperscript{2,3}}

\address{$^1$ Department of Physics, University of Oslo, Oslo, Norway}
\address{$^2$ John Adams Institute for Accelerator Science at University of Oxford, Oxford, UK}
\address{$^3$ Deutsches Elektronen-Synchrotron DESY, Hamburg, Germany}

\ead{c.a.lindstrom@fys.uio.no}

\begin{abstract}
This contribution outlines the HALHF concept, which combines the high gradients achievable in plasma-wakefield acceleration with conventional radio-frequency acceleration. In HALHF, beam-driven plasma-wakefield cells are used to accelerate electrons to high energy. Because plasma-based acceleration of positrons is problematic, conventional RF acceleration is used but to much lower energy. The HALHF concept utilises not only asymmetric energies but also asymmetric bunch charges and asymmetric transverse  emittances, leading to comparable luminosity to conventional facilities but much lower capital cost.  Possible upgrades to the HALHF facility are discussed, in particular to the $\rm{t\bar{t}}$ threshold and to \SI{550}{GeV}, where the Higgs self-coupling and $\rm{t\bar{t}}H$ coupling can be measured. Other upgrades include the provision of two interaction points, to implement a $\gamma$--$\gamma$ collider of two possible types and finally a symmetric high-energy collider if the problem of plasma-based positron acceleration can be solved.  
\end{abstract}

%%%%%%%%%%%%%%%%%%%%%%%%%%%%%%%%%%%%%%%%%%%%%%%%%%%%%%%%%%%%%%%%%%%%%%%%%
\section{Introduction}
\label{sec:intro}
The hybrid, asymmetric, linear Higgs factory (HALHF) and its physics and practical motivation have been described in detail in Ref.~\cite{HALHF}. The concept is founded on utilising the extremely high gradients attainable in plasma-wakefield accelerators (PWFA) \cite{Veksler_HEACC_1956,Tajima_PRL_1979,Chen_PRL_1985,Ruth_PA_1985} while avoiding the problematic plasma-wave acceleration of positrons~\cite{Cao_arXiv_2023}. In this contribution, we briefly outline again the main points of the HALHF concept, together with its estimated cost. We expand on details of possible upgrades to the HALHF concept given in the original paper. Firstly, an upgrade to reach the $\rm{t\bar{t}}$ production threshold is examined and shown to be possible with minimal changes to the layout and cost of HALHF. A further energy upgrade to reach \SI{550}{GeV}, where the Higgs self-coupling and top Yukawa coupling can be measure is also assessed, although more schematically. Next, a variation on the HALHF scheme that naturally provides two interaction points for independent and complementary detectors is outlined. This gives the possibility of enhanced overall luminosity as well as a number of other advantages. Schemes for $\gamma$--$\gamma$ colliders are outlined, which require a new interaction hall in the centre of the HALHF footprint. In particular, should any of the currently proposed~\cite{Vieira_PRL_2014,Diederichs_PRAB_2019,Silva_PRL_2021,Zhou_PRL_2021,Hue_PRR_2021} or future methods of positron acceleration in PWFA prove practical, this upgraded layout has significant advantages; it greatly facilitates a fully PWFA-based, symmetric future incarnation of HALHF.

\section{Asymmetric beam energy, current and emittance}
\label{sec:beam-specs}

The centre-of-mass (c.o.m.) energy for a Higgs factory is conventionally chosen to be \SI{250}{GeV}. Cost minimisation of a hybrid technology, in which PWFA is much cheaper than a conventional RF linac, implies that the length of the RF-based positron arm should be minimised, resulting in an asymmetric-energy collider. The length of the positron linac is determined by an optimisation of efficiency, capital cost, and facility and carbon footprint. We chose a positron energy of \SI{31.3}{GeV}. Simple relativistic kinematics implies that the electron energy required to give the required c.o.m.~energy is a four times higher than for a symmetric machine, {\it viz.}~\SI{500}{GeV}. This produces a boost of $\gamma = 2.13$, smaller than the value of 3 at the HERA electron/positron-proton collider.

The HALHF and HERA colliders are not strictly comparable, since the proton is composite whereas the electron and positron are point-like. This means that the overwhelming majority of collisions at HERA have a effective boost much smaller than 3. Nevertheless, HERA did successfully analyse collisions at a high fraction of the proton's total momentum, $x$. Examples with similar boosts to that at HALHF include limits on heavy-resonance production at threshold, e.g.~for leptoquarks~\cite{HERA_LQ} and high-$x$ structure-function determinations~\cite{HERA_highx}. 
This gives confidence that a similarly successful detector as those at HERA could be designed for HALHF. 

Asymmetric beam energies are less energy efficient. For equal bunch charges, the above beam-energy specification would increase energy usage 
by a factor $\gamma=2.13$ compared to the symmetric case. A charge asymmetry, reducing the charge of the high-energy electron beam by a factor four while increasing that of the positron beam by the same factor produces identical geometric luminosity (ignoring beam–beam effects) for the same power consumption. However, it is difficult to produce such a large number of positrons, so that a more conservative factor of two was chosen for HALHF, which results in an energy efficiency only 25\% less than for a symmetric machine. 

Asymmetric beam energies do not affect the geometric luminosity as long as beam sizes remain constant. However, the ``hour-glass" effect \cite{Furman_SLAC} and the beam--beam effect \cite{Schulte_BB_2017} must be taken into account. By varying the beta functions and bunch lengths appropriately, the normalised emittance of the electron beam, which is accelerated using PWFA, can be increased by a factor of 16 compared to a symmetric machine. Using GUINEA-PIG \cite{Schulte_Thesis_1996}, the resultant HALHF luminosity can be compared to that of the symmetric ILC machine~\cite{ILC_TDR_Accelerator}. The luminosity per bunch crossing decreases by 37\%, while that within 1\% of the peak energy decreases by 50\%. At the $Z^0$, the equivalent reductions are 28\% (50\%).

\section{Schematic layout of the collider}
\label{sec:Setup}

\begin{figure*}[htb]
	\centering\includegraphics[width=\textwidth]{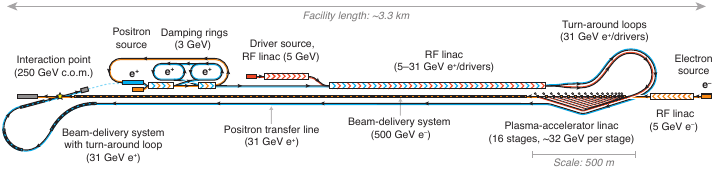}
	\caption{Schematic of HALHF. Particle sources provide electrons (orange), positrons (blue) and electron drivers (red). Electrons are accelerated to \SI{5}{GeV} and diverted via a return loop back to produce the positrons in the positron-source complex. The positrons are captured, accelerated to \SI{3}{GeV} and injected into a pre-damping ring. A second damping ring produces low-emittance bunches that are accelerated in the linac to \SI{31.3}{GeV} before being turned around. The positrons enter the beam-delivery system, which is combined with another turn-around, and then enter the final focus to collisions. Trains of electron drivers are accelerated to \SI{5}{GeV} in a dedicated RF linac, then accelerated with a leading positron bunch to \SI{31.3}{GeV}. The drive beams are separated after turn-around and injected into sixteen PWFA stages, accelerating an electron bunch from a photocathode injector up to \SI{500}{GeV}. The spent electron drivers are steered into separate beam dumps. The accelerated electron bunch enters the high-energy beam-delivery system before colliding with the positron bunch. The spent colliding beams enter beam dumps located after the interaction point. The dashed line represents an option to re-use the spent positrons. From Ref.~\cite{HALHF} (CC BY 4.0).}
    \label{fig:1_setup}
\end{figure*}

Figure~\ref{fig:1_setup} shows the schematic layout of the HALHF facility. There are two principal accelerators~--~one an RF-based linac for the electron drive beams and positrons, the other a plasma-based linac for high-energy electrons~--~and three particle sources: one for electrons to produce positrons, one for the electron drivers, and one for the colliding electrons. Other components include positron damping rings, beam-delivery systems for the colliding beams, and transfer lines. A complete summary of the HALHF parameters used is given in Ref.~\cite{HALHF} and is not repeated here.

%--- FIGURE 1: setup figure ---%

%\renewcommand 
%\thesubsection{\arabic{subsection}}

\subsection{Electron sources}
Beam-quality requirements are low for the drive beams and, to a lesser extent, the colliding electron beam, allowing photocathode injectors to be utilised and obviating the need for a dedicated electron damping ring~\cite{Xu_DR_Free}. The colliding electrons should be highly spin polarised. The colliding-beam source produces bunches of $1\times10^{10}$ electrons (\SI{1.6}{nC}), while the other produces trains of drive-beam bunches, each with $2.7\times10^{10}$ electrons (\SI{4.3}{nC}). A train of 16 drive-beam bunches (one for each stage of the plasma-based linac) is produced for each colliding bunch, separated by \SI{5}{ns}, determined by the rise time of extraction kickers (2--\SI{4}{ns}) and the transit time of the accelerated bunch between PWFA stages.

\subsection{Positron source}
\label{sec:positronsource}
A ``conventional" positron source~\cite{Chalkovska_positrons} such as that proposed for CLIC \cite{CLIC_CDR_2012} is the baseline for HALHF. Around $4\times10^{10}$ positrons (\SI{6.4}{nC}) are required per bunch, which is challenging. The spent \SI{31.3}{GeV} positron beam could also possibly be captured and reused.

\subsection{Main RF linac for positron and drive-beam acceleration}
\label{sec:RFlinac}
The conventional linac accelerates the electron drive bunches for the plasma-based linac and the colliding positrons to \SI{31.3}{GeV}. Its gradient is assumed to be \SI{25}{MV/m}, giving a total length of \SI{1.25}{km}. Technological options considered for the linac are an L-band normal-conducting RF linac or continuous-wave (CW) superconducting linac. The average power, delivered by conventional klystrons, if the linac operates at \SI{100}{Hz} (see Sec.~\ref{sec:bunchtrainpattern}) is \SI{21.4}{MW}.

\subsection{PWFA linac for high-energy electrons}
\label{sec:PWFAlinac}
The drivers are turned around in a loop and distributed by a set of kickers in a tree-like delay chicane \cite{PfingstnerIPAC2016} to synchronize them with the colliding electron beam. 

The plasma-accelerator stages, each \SI{5}{m} long, has a plasma density of $7\times10^{15}$~cm$^{-3}$ and an accelerating gradient of \SI{6.4}{GV/m}, accelerates the electron beam by just under \SI{32}{GeV} per stage to produce a \SI{500}{GeV} beam. The first of the 16 stages is shorter (\SI{2.5}{m}), accelerating the beam from \SI{5}{GeV} to around \SI{20}{GeV}, which is sufficiently different from the \SI{31.3}{GeV} drive beam energy to allow injection into the second stage.  The total length of plasma accelerator is only approximately \SI{80}{m}. The remainder of the \SI{410}{m} total length consists of magnetic chicanes, on average \SI{26}{m} per stage but scaling with the square root of the energy, which exploit a self-correction mechanism in longitudinal phase space \cite{Lindstrom_preprint_2021}. The average accelerating gradient is therefore \SI{1.2}{GV/m}. Depending on the exact assumptions, at least 25\% of the power in the driver train is dumped into each beam dump, corresponding to \SI{375}{kW} per beam dump and a total of \SI{6}{MW}. Half of the remaining power is dumped into the plasma accelerators themselves, corresponding to \SI{93}{kW/m}. Cells that can cope with such energy depositions must be developed.

Preserving the emittance\footnote{Preserving electron polarisation is in principle possible in a plasma accelerator \cite{VieiraPRSTAB2011}, although this has not yet been experimentally verified.} will be a major challenge for the plasma-based linac~\cite{Lindstrom_Thevenet}. Fortunately, the HALHF concept allows normalised emittances to as high as $160 \times 0.56$~mm-mrad, achieving which requires transverse instabilities to be suppressed~\cite{LehePRL2017,Mehrling2018}. Misalignment tolerances must be less than \SI{100}{nm}~\cite{LindstromIPAC2016,Schulte2016RAST} (in the vertical plane). Such constraints are tighter than for CLIC or ILC, but are required over a much shorter distance. 

The performance of the PWFA arm is clearly the most uncertain of all HALHF elements but can be relatively easily modified to cope with problems. Achieving acceptable jitter will be a major aim of future R\&D.

\subsection{Overall bunch-train pattern}
\label{sec:bunchtrainpattern}
The bunch-train pattern depends on the linac technology chosen; a normal-conducting RF linac must be operated in burst mode (with trains of bunches), whereas a superconducting RF linac can in principle be operated CW. The number of drivers per colliding bunch is determined by the number of stages (i.e., 16). Since the required delay between stages and the rise time of available kickers requires that drivers are separated by $\sim$\SI{5}{ns}, the shortest gap between colliding bunches will be \SI{80}{ns}, consistent with the maximum repetition rate tolerable in a plasma~\cite{DArcyNature2022}.  

Approximately 10,000~bunch collisions per second are necessary to compensate for the luminosity loss per bunch from beam--beam and hour-glass effects (Sec.~\ref{sec:beam-specs}), which exceeds that for ILC (6,\SI{560}{Hz}), but is smaller than CLIC (15,\SI{600}{Hz}). 
This requires about \SI{100}{Hz} of bunch trains and thus $\sim$160,000 drivers per second.

\subsection{Damping rings}
\label{sec:dampingrings}
The length of the positron damping rings is $\sim$\SI{300}{m} plus additional gaps for kickers and wigglers, comparable to the CLIC main damping rings, which have a circumference of \SI{359}{m}. Since the HALHF injection energy is \SI{3}{GeV} rather than \SI{2.86}{GeV}, a radius of \SI{400}{m} is assumed. The damping rings have a characteristic damping time of approximately 1--\SI{2}{ms}~--~short enough to fully damp within the \SI{10}{ms} period between bursts. These estimates however are provisional, awaiting a detailed design of the system.

\subsection{Beam-delivery systems}
\label{sec:bds}
The HALHF electron beam-delivery system (BDS) is assumed to be identical to ILC, which was designed for the same energy (\SI{500}{GeV}) and is \SI{2.25}{km} in length. It must also stretch the electron bunch from \SI{18}{\micro\meter} rms to \SI{75}{\micro\meter} rms, which can be done in the long magnetic chicane used for energy collimation. The electron BDS drives the total length of the HALHF facility. That for the positrons is much shorter. Raimondi and Seryi~\cite{Raimondi_PRL_2001} estimate that the final focus scales between $E^{2/5}$ and $E^{7/10}$, depending on assumptions. Other parts of the BDS scale at least proportional to energy~\cite{White_BDS}. For definiteness, we assume a scaling by $\sqrt{E}$, resulting in a length of \SI{0.56}{km}. The scaling uncertainty pending a detailed BDS design effort has negligible effect on the HALHF cost or footprint.

\section{Cost estimate}
\label{sec:cost}
The HALHF capital cost is estimated by scaling from other mature projects, principally ILC and CLIC, which have gone through extensive and detailed expert cost-review procedures. 

\subsection{Capital cost}
\label{sec:cost_capital}
Conventional subsystems and civil construction dominate the cost of HALHF. No attempt has been made to cost the personnel required by HALHF. However, it should be considerably smaller than that required by ILC; scaling by the overall capital costs of the two projects should give a reasonable estimate. The cost of HALHF is $\sim$1.55 billion ILCU\footnote{1 ILCU is defined as \$1 on Jan.~1st, 2012.}, based on the subsystem breakdown shown in Table~\ref{tab:1}. 

The HALHF cost today can be estimated from the USA GDP deflator to be approximately \$1.9B. The Implementation Task Force (ITF) parameterization prepared for the Snowmass 2021 process~\cite{Snowmass_Implementation} gives the Total Project Cost (TPC), sometimes referred to as ``US accounting". Applying this methodology to HALHF gives a TPC of \$4.46B~\cite{Spencer_private}, much smaller than any other Higgs factory proposal.

\begin{table*}[t]
    \begin{tabular}{p{0.33\linewidth}>{\centering}p{0.15\linewidth}>{\centering}p{0.13\linewidth}>{\centering}p{0.15\linewidth}>{\centering}p{0.1\linewidth}}
        \textit{Subsystem} & \textit{Original cost (MILCU)} & \textit{Scaling factor} & \textit{HALHF cost (MILCU)} & \textit{Fraction} \vspace{0.55cm} \cr
        \hline
        Particle sources, damping rings & 430 & 0.5 & 215 & 14\% \cr
        RF linac with klystrons & 548 & 1 & 548 & 35\% \cr
        PWFA linac & 477 & 0.1 & 48 & 3\% \cr
        Transfer lines & 477 & 0.15 & 72 & 5\% \cr
        Electron BDS & 91 & 1 & 91 & 6\% \cr
        Positron BDS & 91 & 0.25 & 23 & 1\%  \cr
        Beam dumps & 67 & 1 & 80 & 5\% \cr
        Civil engineering & 2,055 & 0.21 & 476 & 31\% \cr
        \hline
        Total &  &  & 1,553 & 100\% \cr
    \end{tabular}
   \caption{Estimated capital construction cost of HALHF by subsystem. The costing is based on scaling the estimated costs of the nearest equivalent subsystem of CLIC, ILC or European XFEL; see Ref.~\cite{HALHF} for further details.}
   \label{tab:1}
\end{table*}

\subsection{Running costs}
\label{sec:cost_running}
The running costs are dominated by wall-plug-power usage, which is estimated mostly by analogy to other projects. It is dominated by the drive beams. Assuming 50\% wall-plug efficiency, this requires around \SI{48}{MW}. Damping rings, of which there are two, add $\sim$\SI{10}{MW} each~\cite{Charles2018}. Cooling power requirements are assumed to be similar to CLIC, which adds roughly 50\% of the RF power requirement. This totals $\sim$\SI{92}{MW} rounded up to \SI{100}{MW} by magnet requirements, etc., which is similar to ILC and CLIC Higgs factories.

\section{Upgrade concepts}
Several paths for upgrading HALHF beyond the initial concept can be envisaged, some of which are outlined below. Additional costs are given in European accounting, i.e. refer to those given in Table~\ref{tab:1}. 

\subsection{HALHF with polarised positrons}
\label{sec:polaristion}
Although the original version of HALHF used unpolarised positrons, there are very significant physics advantages to having polarised positrons, even if the degree of polarisation is significantly lower than that of the electrons. If the simulation-based prediction that polarised electrons can be maintained in PWFA turns out to be too optimistic, having the possibility of polarised positrons becomes even more important. An alternative scheme is to adapt the ILC scheme. The spent high-energy electron beam could be passed through a wiggler, the resultant photon beam striking a rotating target~\cite{ILC_TDR_Accelerator}. Although there are still some outstanding problems with the ILC design, including the rotating target, these could likely be solved if enough engineering design effort were available. The application at HALHF is also more challenging than at ILC because of the higher energy of the electron beam, which would require a longer wiggler and other modifications to the ILC design. On the other hand, this method of generating positrons would save significant power. The indicative cost of the ILC positron generation is \SI{300}{MILCU}; taking into account savings on the conventional positron source, the additional cost of this rather straightforward upgrade would be $\sim$\SI{185}{MILCU}, 12\% of the original HALHF cost. 

\subsection{HALHF at the $\rm{t\bar{t}}$ threshold (380 GeV c.o.m.)}
\label{sec:ttbar}

\begin{figure*}[t]
	\centering\includegraphics[width=\textwidth]{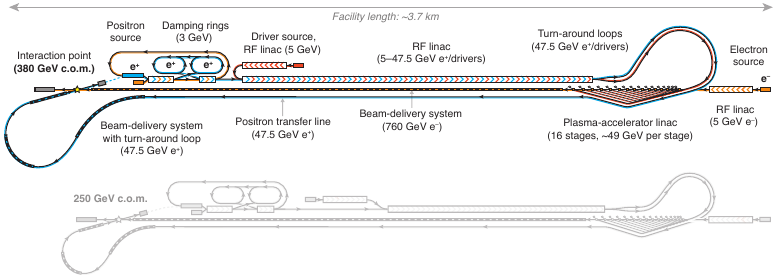}
	\caption{Schematic layout of the \SI{380}{GeV}-option for the hybrid asymmetric linear Higgs factory, operating at the $\rm{t\bar{t}}$ threshold. The overall facility length increases by approximately 10\%. A greyed-out version of the original \SI{250}{GeV}-option is shown below for comparison.}
    \label{fig:2_ttbar}
\end{figure*}

The c.o.m.~energy threshold for $\rm{t\bar{t}}$ production is $\sim$\SI{346}{GeV}; to allow for the top-quark width of $\sim$\SI{2}{GeV}, running at \SI{350}{GeV} would be sufficient. However, a slightly higher energy allowing the observation of the $WW$ fusion process allows the total top-quark width to be measured accurately. A choice of \SI{380}{GeV} is therefore usually made. If the positron energy of \SI{31.3}{GeV} is retained, then the required electron energy is $\sim$\SI{1165}{GeV}. The relativistic $\gamma$ factor is 3.15, still similar to that of HERA. The spherical nature of top-quark decays in its c.o.m.~means that despite the boost, most of the decay products will remain in the acceptance of the detectors. However, the energy inefficiency in such a configuration would be increased by 35\% compared to Higgs running. If more positrons could be produced (i.e., using a higher charge asymmetry), this loss in efficiency could be regained, but the production of positrons itself is energy intensive and very technically difficult. Required modifications to the original HALHF concept would involve an adaption of the electron BDS to cope with the more than doubling of the electron energy. As remarked earlier, the various parts of the BDS scale differently with energy; however, it is instructive to note that the ILC BDS length at \SI{500}{GeV} beam energy is \SI{2.25}{\kilo\metre} long, whereas that for CLIC, which is designed for beam energies of \SI{1.5}{TeV}, is only \SI{2.75}{\kilo\metre}. The fact that these two lengths are comparable is a function of the assumptions made by the two designs and the degree of risk and background levels considered acceptable~\cite{White_BDS}. Given that we require a beam energy smaller than \SI{1.5}{TeV}, we make the perhaps optimistic assumption that the ILC BDS length will suffice for $\rm{t\bar{t}}$ operation, also assuming of course appropriate replacement of components, increase of magnet currents etc. The price for this assumption might well be a requirement to tolerate higher backgrounds in the detectors.  % TODO

%With the above assumptions, \textbf{the only other changes required} to the original HALHF concept would be an increase in the number of cells (from 16 to 37), which increases proportional to the electron energy and in the required average interstage matching length, which increases as the square root of the beam energy. The total length required is \SI{1.42}{\kilo\metre}, an extra length compared to the Higgs-factory of order \SI{1}{\kilo\metre}, which would increase the HALHF footprint substantially. The extra costs that would affect the Higgs-factory HALHF are the increased tunnel length of \SI{48}{MILCU}, which would initially be left empty. To implement $\rm{t\bar{t}}$ operation, additional beam dumps for the drive beams (\SI{105}{MILCU}) and the extra instrumentation required for the additional 21 PWFA stages and interstage optics (\SI{63}{MILCU}) would be required. This totals \SI{213}{MILCU}, $\sim$14\% of the original HALHF capital cost. 

However, the option of increasing the electron energy while leaving the positron energy unchanged has significant disadvantages. In particular, the number of stages must be more than doubled, requiring the generation of an equivalent number of additional drive beams. This would be impractical in an already heavily loaded linac.

An alternative configuration is to keep $\gamma$ constant by increasing the energy of both beams. A c.o.m.~energy of \SI{380}{GeV} and $\gamma$ of 2.13 implies $E_p = \SI{47.5}{GeV}$ and $E_e = \SI{760}{GeV}$. For the positrons, either the gradient of the linac could be increased appropriately, or, more likely given the already high loading, the linac could be lengthened by $\sim$50\% to \SI{1.875}{\kilo\metre}. This could best be managed by increasing the lengths of the footprint and the appropriate tunnels of the initial Higgs-factory configuration and either leaving them empty for Higgs running or installing the full components e.g.~the linac and running it at a lower gradient. The latter option would increase the capital cost for the Higgs-factory HALHF project by $\sim$\SI{270}{MILCU}, the former by \SI{58}{MILCU}\footnote{These numbers are very approximate. For the ``empty linac extention" option, they were obtained by scaling the civil construction cost by the ratio of footprint lengths. For the ``full linac" option, they were obtained by adding the extra civil cost to the scaled cost of the linac, minus the cost of klystrons which would not be installed for Higgs running. Klystrons of \SI{300}{kW} costing \SI{2}{MILCU} each were assumed.} For the electrons, the increase in energy of the drive beams to \SI{47.5}{GeV} by the linac modification would allow the required energy increase without increasing the number of stages, although the length of each plasma cell would increase proportionately (to \SI{7.6}{m} per stage). The required average interstage length also increases, giving an extra length requirement for the PWFA arm of $\sim$\SI{120}{\metre} (\SI{530}{\metre} total). As discussed above, the BDS would be left unchanged and assumed to be able to cope with the increase in electron energy, at the risk of higher detector backgrounds. A proportional increase in the turn-around radius, the lengthening of the PWFA arm and slight increases elsewhere to cope with end effects increase the overall footprint slightly to \SI{3.7}{\kilo\metre}. The estimate for the overall cost increase of a $\rm{t\bar{t}}$ HALHF compared to the original HALHF Higgs factory is therefore $\sim$\SI{350}{MILCU}. Running costs would increase because of the additional linac power requirements and cooling by \SI{36}{MW} and including increased magnet current at higher energy gives a total of approximately \SI{140}{MW}.

In conclusion, the option of increasing the energy of both beams to keep $\gamma$ constant is preferable. The new HALHF layout showing the additional footprint length for the $\rm{t\bar{t}}$ upgrade is shown in Figure~\ref{fig:2_ttbar}.

\subsection{HALHF and Higgs couplings (550 GeV c.o.m.)}
\label{sec:550GeV}

A crucial measurement to elucidate Higgs physics is the value of the Higgs self-coupling. Studies for linear colliders indicate that a minimum c.o.m.~energy of \SI{500}{GeV} is required~\cite{DiMicco}. Increasing this to \SI{550}{GeV} would greatly increase the cross section for the $\rm{t\bar{t}}H$ process and improve the accuracy on the top Yukawa coupling by a factor two\cite{Barklow-ILC}. Increasing the HALHF energy to  \SI{550}{GeV} would follow the same procedure as outlined in the previous section. Keeping the boost approximately constant, the positron energy would be increased to \SI{68.5}{GeV} and the electron energy to \SI{1.1}{TeV}. This would result in a lengthening of the RF linac by more than a factor of two compared to the Higgs Factory (i.e., the \SI{250}{GeV} option), to \SI{2.75}{km}. This is now sufficiently long that it would become a driver of the HALHF footprint, which would grow beyond the \SI{3.7}{km} of the $\rm{t\bar{t}}$ upgrade. The length of the plasma-acceleration arm would also increase. Assuming the plasma accelerating gradient is unchanged, 16 cells each of \SI{11}{m} would be required. Incorporating the increase in the interstage matching optics would result in a total plasma-arm length of $\sim$\SI{664}{\metre}, an increase of \SI{254}{\metre} compared to the Higgs factory\footnote{For such a large energy increase, a re-optimisation with respect to beam energy, number of stages, plasma gradient etc.~should be carried out.}. The additional cost is dominated by the \SI{650}{MILCU} for the linac. There would in addition be extra tunneling costs which in the absence of re-optimisation of the parameters are difficult to estimate. A reasonable guess for the extra cost of this option would be to round up to \SI{750}{MILCU}, a roughly 48\% increase over the cost of the Higgs factory. The running costs would increase by approximately \SI{90}{MW} in the linac plus cooling, giving a total power of around \SI{190}{MW}.

\subsection{Two HALHFs: HALHF with two interaction points}
\label{sec:twoip}

\begin{figure*}[b]
	\centering\includegraphics[width=\textwidth]{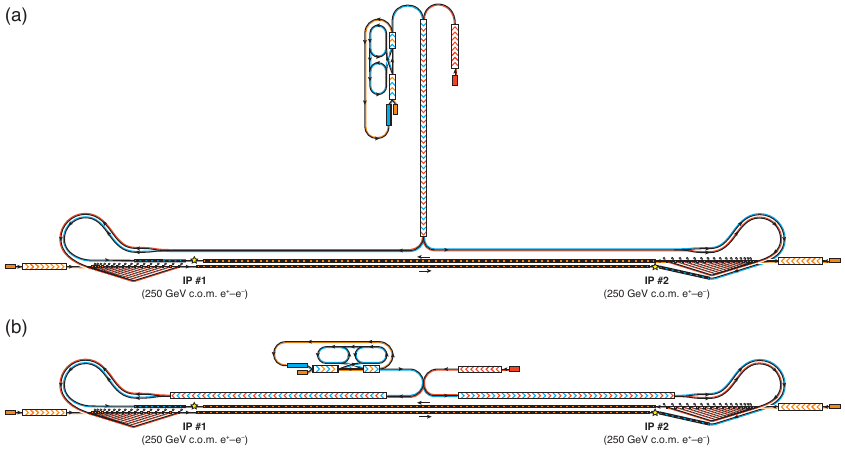}
	\caption{Schematic layout of the hybrid asymmetric linear Higgs factory with two interaction points, with either (a) a single RF linac in a T-shaped configuration, or (b) with two separate RF linacs for increased beam power.}
    \label{fig:3_Tshape}
\end{figure*}

One of the acknowledged drawbacks with a linear collider is that, since the colliding beams interact only once, there is no integrated luminosity gain in having two detectors, since the luminosity can only be shared. This is in contrast to a circular collider, where additional interaction points with approximately identical instantaneous luminosity can be added around the ring. There are significant advantages to having at least two detectors at a Higgs factory, such as allowing complementary technologies to be used, heightening sensitivity to particular phenomena and increasing discovery potential. However, a cost-benefit analysis of two detectors at a linear collider such as ILC is ambiguous. It was decided not to build two separate interaction points with the concomitant separate and expensive BDS infrastructures. The ILC solution was to construct two detectors and move them alternatively in and out of the beam line. The frequency of this was not decided at the time of the TDR but clearly for reasons of efficiency could not be more frequent than once or twice a year. The CLIC CDR, in contrast, only considered one detector at the interaction point.

Neither of the above solutions is ideal. For HALHF, there are possibilities that are not only more tractable than ILC or CLIC, but also offer advantages when considering future upgrades (see Secs.~\ref{sec:gammagamma} and \ref{sec:pwfapositrons}). The linac in the original HALHF scheme would be rotated by \SI{90}{\degree}, so that the drivers and positron bunches destined for each IP are kicked to different sides on leaving the linac and then transported to the appropriate interaction point. Although the additional tunnel construction required for such a configuration is relatively small, the footprint of the facility is larger and its ``T" shape is less easy to fit on existing laboratory sites. This is illustrated in Fig.~\ref{fig:3_Tshape}(a).

The cost of providing two interaction points in these schemes is a duplication of the BDS, the excavation of a second detector cavern, the provision of a second PWFA arm to source and accelerate electrons to \SI{500}{GeV} and a variety of transport lines. However, there are also some savings: the long positron transfer line (including the turnaround in the positron BDS) in the original HALHF configuration no longer exists as the IPs have effectively been moved closer to the end of the linac. Since the two electron BDSs could be placed parallel in the same tunnel (which would be somewhat larger than considered for the original HALHF configuration), the extra cost would only be that of the required beamline elements, such as magnets, collimators etc. This extra cost can be seen from Table~\ref{tab:1} to be \SI{91}{MILCU}. The extra tunnel for the linac would cost an additional \SI{60}{MILCU}. Other costs include \SI{48}{MILCU} for the duplicate PWFA linac and $\sim$\SI{100}{MILCU} for an additional electron source. The cost of the second interaction hall is not considered here, this being to some extent offset by the loss of the long positron transport line. The total extra costs for this 2-IP option is $\sim$\SI{300}{MILCU}, a roughly 19\% increase on the original HALHF cost.

A final option, which is considerably more expensive but would have other advantages such as higher overall luminosity, is to build a second linac to feed the second IP. This would be constructed in the same tunnel that houses the original HALHF linac, which would be doubled in length~--~the layout is shown in Fig.~\ref{fig:3_Tshape}(b). The total luminosity summed over the two experiments could be doubled from the original HALHF design, but only if sufficient positrons could be produced. This might require the construction of a second positron source. While the footprint is smaller than that of the ``T"-shape, this scheme is considerably more expensive, since the additional linac alone would correspond to 35\% of the original HALHF cost. The overall increase in cost of this solution is approximately \SI{689}{MILCU}, 44\% larger than the original HALHF cost; an additional positron source would require another $\sim$\SI{115}{MILCU}, corresponding to an increase of 52\%.

Given the advantages of a two-IP system mentioned above, one of these upgrade variants could well be desirable. Such an upgrade would allow the possibility of one detector continuing to take data while the other may be being upgraded, or require repairs. In such a situation, no luminosity would be lost as all colliding bunches could be routed though one of the detectors, exactly as in the original HALHF scheme. 
%In addition, sociological effects such as the friendly rivalry between two competing groups of physicists are important. This adds both a sense of urgency and the imperative for careful checking of all published results. 

\subsection{Two HALHFs: $\gamma$--$\gamma$ colliders}
\label{sec:gammagamma}

Linking a $\gamma$--$\gamma$ collider with an $e^+$--$e^-$ collider has a long history \cite{Ginzburg83,Ginzburg84,Telnov}. Recently, this idea has been revisited in inputs to the US Snowmass 2021 process \cite{Barklow, Barzi}. Similar configurations can be envisaged for an upgraded HALHF facility. The idea is to use a high-powered laser to Compton scatter from the electron or positron beam, which produces a flux of photons with a peak energy around 83\% of that of the scattering charged-particle beam and highly collimated in this beam’s initial direction. 

Starting with the 2-IP HALHF facility described in Sec.~\ref{sec:twoip}, relatively modest changes would be required to produce a symmetric $\gamma$--$\gamma$ collider similar to that traditionally proposed. In particular, a third interaction point specially optimised for laser back-scattering on the electron beams would need to be constructed at the centre of the facility to intersect the two BDS systems. For $\gamma$--$\gamma$ operation, the positron source would be switched off and only electron bunches employed. A wide c.o.m.~energy range can be explored; from the \SI{10}{GeV}-scale to \SI{1}{TeV} (or \SI{2.2}{TeV} if the PWFA linacs are upgraded as discussed in Sec.~\ref{sec:550GeV}). A $\gamma$--$\gamma$ Higgs factory could be achieved by using beams at \SI{86.5}{GeV} \cite{Barklow} (using the $\gamma\gamma \to H$ channel), requiring only three PWFA cells and associated drive beams. The electron beams would be collided with suitable laser beams to produce Compton-scattered photons for the $\gamma$--$\gamma$ collisions.

\begin{figure*}[t]
	\centering\includegraphics[width=\textwidth]{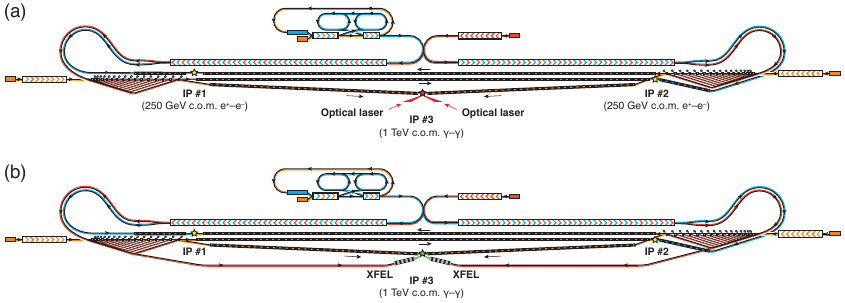}
	\caption{Schematic layout of the HALHF facility as a $\gamma$--$\gamma$ collider, which has a new IP at the center. The back-scattered photon beam can be produced either with (a) optical lasers or (b) X-ray FELs.}
    \label{fig:4_gamma-gamma}
\end{figure*} 

An additional requirement for a $\gamma$--$\gamma$ upgrade, however, is that while asymmetric $e^+$--$e^-$ collisions can operate with increased emittance (as discussed in Sec.~\ref{sec:beam-specs}), that is not the case for the symmetric $\gamma$--$\gamma$ collisions. This may have implications for the electron sources: while round beams can in principle be used (as photons see no beam--beam effects), the lower emittance may require electron damping rings. The positron damping rings can potentially be repurposed to instead operate with electrons. Furthermore, all the PWFA stages will need to preserve the reduced emittance, which implies stricter tolerances on misalignment and matching etc. It is currently unclear whether such low emittances can be maintained in the PWFA arm: this is a topic of active research.

The cost of such a facility would be relatively small; the BDS would be significantly shorter but a tunnel roughly equivalent to one HALHF BDS would be required to bring the two electron beams to a central IP. This would be the major additional cost, plus the excavation of a suitable experimental cavern and the provision of an appropriate laser. Since, at the time of writing, lasers with the characteristics required for the above facility do not exist, costing such a device is impossible, but is likely to be significantly less than the cost of the civil engineering of the BDS tunnel and its instrumentation, which is $\sim$\SI{200}{MILCU}. A conservative total cost estimate would be \SI{250}{MILCU}.

Much higher $\gamma$--$\gamma$ collision c.o.m.~energies could also be achieved by using electron beams up to the HALHF Higgs-factory maximum of \SI{500}{GeV}, or up to \SI{2.2}{TeV} if the \SI{550}{GeV} upgrade has already been implemented. The HALHF-$\gamma$--$\gamma$ facility is shown schematically in Fig.~\ref{fig:4_gamma-gamma}. Assuming that the increased background is tolerable and fully populating magnetic and focusing elements in the BDS, its length could remain unchanged. In principle, $\gamma$--$\gamma$ collisions could be operated in parallel with $e^+$--$e^-$ collisions in the two regular IPs (each operating at reduced frequency).

Alternative laser technologies can also be considered. One option is to adopt the XCC proposal from Barklow \textit{et al.}~\cite{Barklow}, which uses X-ray beams from two X-ray free-electron lasers (XFELs), but replacing the $\rm{C}^3$ technology\footnote{It would also be possible in principle to utilise $\rm{C}^3$ technology in the HALHF linac; this would require further study.}\cite{CCC} with PWFA. The photon beams are produced in XFELs into which \SI{31}{GeV} electron beams are run. Since this happens to be almost exactly the drive-beam energy of the HALHF Higgs factory, it may be possible to utilise the drive beams directly from the linac, although the drive-beam electron injector would need to be upgraded to produce a high degree of polarisation, $\sim$90\%.

\subsection{Potential for plasma acceleration of positrons}
\label{sec:pwfapositrons}

The culmination of the potentially multi-decadal upgrade programme for the HALHF concept would be the return to a symmetric $e^+$--$e^-$ collider. This would be possible if by then the problem of plasma acceleration of positrons had been solved. In this case, if a two-IP or $\gamma$--$\gamma$ collider facility already existed, the switch to a symmetric, plasma-based $e^+$--$e^-$ collider can easily be envisaged -- especially if positron PWFA can be driven by electrons. It appears that no major infrastructure changes would then be required beyond those needed for the $\gamma$--$\gamma$ collider; the limitation is technological rather than financial. While it is difficult to cost an upgrade based on currently unknown technology, a reasonable assumption would be that the cost of a positron-PWFA arm would be identical to that of the electron arm. In this case, no extra cost would be required for a symmetric machine, although as the maximum c.o.m.~energy is increased, changes to the facility would certainly be required. Such a symmetric machine would in principle give access to multi-TeV collisions.

%% CONCLUSIONS

\begin{table*}[h]
    \begin{tabular}{p{0.51\linewidth}>{\centering}p{0.18\linewidth}>{\centering}p{0.23\linewidth}}
        \textit{Upgrade} & \textit{Additional cost (MILCU)} & \textit{Fraction of original HALHF cost} \vspace{0.05cm} \cr
        \hline
        Polarised positrons & 185 & 12\% \cr
        $\rm{t\bar{t}}$ threshold (380 GeV c.o.m.) & 350 & 23\% \cr
        Higgs self-coupling (550 GeV c.o.m.) & 750 & 48\% \cr
        Two IPs & 300 & 19\% \cr
        Two IPs + additional linac & 689  & 44\% \cr
        Two IPs + additional linac \& positron source & 804  & 52\% \cr
        $\gamma$--$\gamma$ collider (laser-based) & 250 & 17\% \cr
        $e^+$--$e^-$ collider, symmetric (assuming $e^+$ PWFA) & $\sim$0 & $\sim$0 \cr
        \hline
    \end{tabular}
   \caption{Estimated cost of upgrades discussed in the text. The final two upgrades require the ``Two IPs + additional linac \& positron source" upgrade to have already been carried out.}
   \label{tab:2}
\end{table*}

\section{Toward an optimal parameter set}

The initial parameter set for HALHF was intended to be a starting point for an iterative process of refinement and optimisation to reach a self-consistent optimal set for a pre-conceptual design report (pre-CDR). This process has now begun, with task forces beginning work in several areas of the HALHF concept. In particular, the parameters of the plasma acceleration arm are being revised to address a number of questions, including transverse instabilities, ion motion, field ionisation and the effects of betatron radiation reaction (as well as the interplay between these). The linac design, positron generation, preservation of polarisation inside the plasma-accelerator stages and the beam delivery system have been formed and have commenced work. Other task forces, e.g.~damping rings, are in the process of formation. Consideration of compatibility with some of the upgrades described in the previous section may also result in parameter changes. The aim is to update the parameters of HALHF by the end of 2024.

\section{Summary \& Conclusions}
\label{sec:Conclusions}
The HALHF collider concept utilises the strengths of PWFA, its high accelerating gradients, while avoiding one of its principal weaknesses, the difficulty in accelerating positrons. It not only promises a much smaller, cheaper and greener Higgs factory, but can also be readily and relatively cheaply upgraded both in energy and in function. Table~\ref{tab:2} shows a summary of the upgrades discussed above, together with estimates of their cost obtained by scaling from the well understood costs of, in particular, the ILC and CLIC projects. The implementation of all these upgrades would constitute a multi-decade programme of running that would represent an exciting future for particle physics. If positron acceleration in plasmas can be harnessed successfully for collider applications, a symmetrical collider based on HALHF would provide a path to multi-TeV lepton collisions.

%%%%%%%%%%%%%%%%%%%%%%%%%%%%%%%%%%%%%%%%%%%%%%%%%%%%%%%%%%%%%%%%%%%%%%%%%
\section*{Acknowledgements}
BF is grateful to the Leverhulme Trust for the award of an Emeritus Fellowship. 
This work was supported by the Research Council of Norway (NFR Grant No.~313770).

%%%%%%%%%%%%%%%%%%%%%%%%%%%%%%%%%%%%%%%%%%%%%%%%%%%%%%%%%%%%%%%%%%%%%%%%%

\end{document}